\documentclass[twocolumn]{aastex631}

\shorttitle{The Dependence of Earth Milankovitch Cycles on Martian Mass}
\shortauthors{Stephen R. Kane et al.}

\begin{document}

\title{The Dependence of Earth Milankovitch Cycles on Martian Mass}

\author[0000-0002-7084-0529]{Stephen R. Kane}
\affiliation{Department of Earth and Planetary Sciences, University of California, Riverside, CA 92521, USA}
\email{skane@ucr.edu}

\author[0000-0001-9800-6723]{Pam Vervoort}
\affiliation{School of Geography, Earth and Environmental Science, University of Birmingham, Birmingham, B15 2TT, UK}

\author[0000-0002-1160-7970]{Jonathan Horner}
\affiliation{Centre for Astrophysics, University of Southern Queensland, Toowoomba, QLD 4350, Australia}


\begin{abstract}

The Milankovitch cycles of Earth result from gravitational interactions with other bodies in the Solar System. These interactions lead to slow changes in the orbit and angular momentum vector of Earth, and correspondingly influence Earth's climate evolution. Several studies have shown that Mars may play a significant role in these Milankovitch cycles, such as the 2.4~Myr eccentricity cycle related to perihelion precession dynamics. Here we provide the results of a detailed dynamical analysis that explores the Earth Milankovitch cycles as a function of the Martian mass to quantify the extent that Mars influences variations in Earth's orbital eccentricity, the longitude of perihelion, the longitude of the ascending node, and obliquity (axial tilt). Our results show that, although the 405~kyr long-eccentricity metronome driven by $g_2$ (Venus) and $g_5$ (Jupiter) persists at all Mars masses, the $\sim$100~kyr short-eccentricity bands driven by $g_4$ (Mars) lengthen and gain power as Mars becomes more massive, consistent with enhanced coupling among inner-planet $g$-modes. The 2.4~Myr grand cycle is absent when Mars approaches zero mass, reflecting the movement of $g_4$ with the Martian mass. Meanwhile, Earth's obliquity cycles driven by $s_3$ (Earth) and $s_4$ (Mars) lengthen from the canonical $\sim$41~kyr with increasing Mars mass, relocating to a dominant 45--55~kyr band when the mass of Mars is an order of magnitude larger than its present value. These results establish how Mars' mass controls the architecture of Earth's climate-forcing spectrum and that the Milankovitch spectrum of an Earth-like planet is a sensitive, interpretable probe of its planetary neighborhood.

\end{abstract}

\keywords{astrobiology -- planetary systems -- planets and satellites: dynamical evolution and stability -- planets and satellites: individual (Mars)}


\section{Introduction}
\label{intro}

Studying the Solar System's orbital dynamics delivers a uniquely high-fidelity laboratory for planetary science, allowing (for example) validation of secular theory, resonance-driven chaos, and long-term stability \citep{laskar1990,quinn1991}. These tested models may then be exported to interpret and characterize the vastly diverse planetary architectures now known around other stars \citep[e.g.][]{robertson2012b,ford2014,winn2015,horner2025}. The bidirectional synergy of using in situ Solar System knowledge to inform exoplanet models, and using exoplanet demographics to stress-test dynamical theory, has become central to questions of formation, evolution, and habitability \citep[e.g.][]{horner2008a,horner2010e,raymond2009c,martin2015b,horner2020b,kane2021d}. In particular, the coupling between system architecture and climate forcing arise naturally from multi-planet interactions and can pace surface conditions on terrestrial worlds both in our system and beyond \citep{deitrick2018a,deitrick2018b,vervoort2022}. Even within the Solar System, the amplitudes and timescales of Earth's orbital evolution are sensitive to the masses and orbits of neighboring planets, underscoring how small architectural tweaks can move climates into different dynamical regimes \citep{lissauer2001c,horner2020a,kane2023a}. The same physics scales to exoplanetary systems, where architectures range from tightly packed chains to widely spaced giants, offering a comparative framework to identify which dynamical–climate linkages are generic versus contingent \citep{winn2015}. Against this backdrop, mapping how specific architectural degrees of freedom modify orbital evolution advances both Solar System science and exoplanet habitability assessments, and complements broader efforts to quantify long-term stability and external perturbations \citep{zink2020c,raymond2024a}.

Earth’s Quaternary climate variability is paced by variations in the planet’s orbital and rotational parameters, collectively known as the Milankovitch cycles \citep{hays1976,berger1978c}. These cycles arise from three principal components: changes in orbital eccentricity, variations in axial obliquity, and the precession of the equinoxes---all modulating the seasonal and latitudinal distribution of solar insolation \citep{laskar2004c,laskar2011a}. This external forcing has driven the waxing and waning of ice sheets since the mid Cenozoic, caused periodic global warming events in the early Cenozoic, and was possibly responsible for multiple major paleoclimate transitions throughout Earth's history \citep[e.g.][]{imbrie1992,edvardsson2002,lisiecki2005,zachos2010,leandro2022,zhang2022e}.

In the Laplace-Lagrange solution, the eccentricities and inclinations of planetary orbits are treated as a system of coupled harmonic oscillators with secular eigenfrequencies $g_i$ and $s_i$ \citep{laskar1990}. The $g_i$ modes determine the evolution of the orbital eccentricities and their perihelion locations, whereas the $s_i$ modes relate to the evolution of the orbital inclinations and their nodes. For example, in the Sun-Jupiter-Saturn system, there are two main $g$ eigenfrequencies, one dominated by Jupiter's motion ($g_5$) and the other by Saturn's ($g_6$). The eccentricity spectrum for Earth's orbit contains power at $\sim$95--125~kyr as well as a highly stable 405~kyr component, the latter tied to the secular frequency difference between Venus and Jupiter, $g_{2}-g_{5}$ \citep{laskar2004c,olsen2019,zeebe2024a} while Earth--Mars coupling plays a central role in modulating the amplitude of the shorter-period $\sim$100~kyr eccentricity cycles. The obliquity cycle, with a characteristic period of $\sim$41~kyr, is driven by the nodal precession of Earth's orbital ellipse and lunisolar torques; the presence of Earth’s large moon stabilizes the spin axis against chaotic excursions \citep{laskar1993b,lissauer2012a}. The climatic precession band ($\sim$19--23~kyr) results from the interaction of axial precession with the nodal precession of Earth’s orbit, with an amplitude that scales with orbital eccentricity \citep{berger1978c}.

While the Moon exerts the dominant control over Earth’s obliquity stability, the modulation of Earth’s eccentricity and precession is strongly influenced by gravitational perturbations from the other terrestrial planets, most notably Mars. In particular, the $\sim$2.4~Myr ``grand eccentricity cycle'' is associated with the difference in the perihelion-precession rates of Earth and Mars ($g_{4}-g_{3}$), while a companion $\sim$1.2~Myr cycle is tied to the difference in nodal precession frequencies ($s_{4}-s_{3}$) \citep{laskar2004c,dutkiewicz2024}. The near-resonant 2:1 relationship between these beats is a recognized driver of chaotic diffusion in the inner Solar System \citep{laskar1990}. Geological records resolve these grand cycles in detail \citep[e.g.][]{ma2019c,zeebe2019b,ikeda2020a,liu2020c,wu2023e,zhang2023k}. Stratigraphic sequences spanning the Mesozoic and Cenozoic also reveal robust imprints of the 405~kyr ``metronome'' that arise from strong nonlinear responses of Earth system processes to precession forcing \citep{vervoort2024}. These are further modulated by the $\sim$2.4~Myr beat so, together, these cycles provide independent constraints on Solar System chaos and deep-time orbital architecture, referred to as a ``geological orrery'' \citep{olsen2019}. These observations demonstrate that the long-period components of the Milankovitch spectrum are sensitive probes of planetary dynamics, including the secular interactions of Mars and the Earth.

Despite the recognized importance of the Earth-Mars orbital coupling, the specific dependence of Earth’s Milankovitch spectrum on the physical properties of Mars, particularly its mass, has not been systematically investigated. Mars may have been a stranded planetary embryo that formed rapidly and was then starved of additional material, accreting most of its mass within the first $\sim$5--10~Myr \citep{dauphas2011b,kleine2009}. Other proposed causes of Mars' relatively low mass include the idea that the disk that formed the terrestrial planets was truncated to a narrow annulus inside $\sim$1~AU \citep{hansen2009c}, or the possibility that Jupiter's early gas-driven migration first displaced material inward and then reversed to sculpt a depleted Mars region and an excited, mixed asteroid belt \citep[e.g.,][]{walsh2011c,pierens2011b,morbidelli2007d,masset2001a}. A complementary model suggests an early outer-planet instability may have perturbed and depleted embryos near Mars' orbit, stunting its growth \citep{clement2018,clement2019b}. Models that include pebble accretion show that Jupiter’s rapid growth to the pebble-isolation mass and its migration can modulate the inward pebble flux and redistribute solids, naturally producing a mass deficiency beyond 1~AU while still allowing efficient growth near Earth and Venus \citep{lambrechts2014b,bitsch2015b,ida2016a,morbidelli2015a}. Finally, there is the possibility that Mars was victim to a giant collision late in its formation, stripping significant mass from the planet \citep[e.g.,][]{andrewshanna2008a}. Together, these geochemical and dynamical constraints explain Mars’ present mass as an outcome of early timing (fast embryo formation) and architecture (giant-planet growth and migration) that limited its long-term feeding zone. 

Linear secular theory predicts that the fundamental secular frequencies $g_i$ and $s_i$ scale with planetary masses and semi-major axis ratios \citep{murray1999a}. Consequently, variations in the mass of Mars are expected to alter the Martian proper frequencies ($g_4, s_4$) and hence the Earth-Mars beat periods (e.g. $g_{4}-g_{3}$, $s_{4}-s_{3}$), potentially modifying both the amplitude and stability of Earth’s orbital cycles \citep{brasser2009}. Studying the influence of a Mars-like planet on Earth's dynamics improves understanding of how planetary formation processes may ultimately affect the architecture of exoplanetary systems and the orbital characteristics of individual planets. 

In this paper we quantify the sensitivity of Earth’s Milankovitch cycles to the mass of Mars using a suite of direct $N$-body integrations. We vary the Martian mass from 0\% to 1000\% of its present value and track the resulting changes in periods and amplitudes of Earth’s Milankovitch cycles across parameter space.  Section~\ref{methods} describes the methodology in the construction of the orbit and obliquity simulations. The results of the simulations are presented in Section~\ref{results}, evaluating the variations in longitude of perihelion, eccentricity, longitude of the ascending node, and axial obliquity. Section~\ref{discussion} provides a discussion of the dynamical interpretation for the derived variations in Milankovitch cycles, and the implications of our results for the architectures and potential climates within exoplanetary systems. Section~\ref{conclusions} imparts concluding remarks and offers suggestions for further work on this topic.


\section{Methodology}
\label{methods}

Here we describe the setup of our simulations and the methodology that is used to carry them out.


\subsection{Dynamical Simulations}
\label{sims}

To carry out our dynamical N-body simulations, we utilized the Mercury Integrator Package \citep{chambers1999} with a hybrid symplectic/Bulirsch-Stoer integrator and a Jacobi coordinate system \citep{wisdom1991,wisdom2006b}. Additionally, we incorporated first-order post-Newtonian relativistic corrections to allow the orbital behavior of the innermost planets to be modeled accurately \citep{gilmore2008}. The methodology described here follows the prescription that has previously been successfully applied \citep{horner2020a,kane2020e,vervoort2022}. The orbital properties of the Solar System planets were extracted from the Jet Propulsion Laboratory (JPL) Planetary and Lunar Ephemerides DE440 and DE441 \citep{park2021}.

The motivation for this study is to assess the influence of Mars on the Milankovitch cycles of Earth. Thus, we conducted a suite of simulations that include all eight planets with initial orbital elements set to the present day canonical values. In our simulations, the mass of Mars was varied from 0\% to 200\% of its present value, in 10\% increments. We conducted an additional simulation in which the mass of Mars was increased to 1000\% of its present value (slightly more massive than Earth), resulting in a total of 22 simulations. The mass for the Earth in the N-body simulations was set to be the combined mass of our planet plus the Moon, whilst the masses of the other six planets (Mercury, Venus, Jupiter, Saturn, Uranus and Neptune) were set to their canonical values. Each simulation was run for $10^8$~years with an integration time step of 1 day and orbital elements recorded every $10^3$~years. These simulations provided a series of orbital parameter evolution results for all 8 planets and, for each Mars mass scenario, with sufficient time resolution to properly capture the primary planet-planet interactions that drive Earth's Milankovitch cycles.


\subsection{Obliquity Calculations}
\label{obliquity}

The orbital parameters of Earth generated by the dynamical N-body simulations described in Section~\ref{sims} were used to calculate obliquity dynamics. Variations in Earth's eccentricity, nodal precession, and perihelion precession affect its axial tilt and orientation through the influence of solar and lunar torques on Earth's equatorial bulge. We applied the spin dynamics model from \citet{laskar1993b}, based on the equations of rigid Earth theory \citep{kinoshita1977a}, following the approach of \citet{vervoort2022}. The parameters that characterize Earth were set to their modern values. Earth completes one full rotation per 24~hours ($\nu=2\pi$~24$h^{-1}$), Earth's axis precesses with period of $25.7\times10^3$~years ($\psi=50.46772\arcsec~yr^{-1}$), the dynamical ellipticity was set to its canonical value ($E_D=0.00328$), and the tidal effect of our moon at its present mass was taken into account, assuming a circular lunar orbit. The initial obliquity of Earth was set to 23.4\degr (see \citet{vervoort2022} for a sensitivity analysis of the obliquity model to initial conditions). The obliquity calculations were performed for each of the 22 N-body simulations over the full $10^8$~year integration time in order to detect the multi-million-year-long obliquity cycles with spectral analysis methods.


\section{Results}
\label{results}

The variations in the Mars mass applied in our simulations (see Section~\ref{sims}) were used to assess the amplitude and periodicity of four major Earth dynamical components: longitude of perihelion ($\varpi$, $g_i$ frequencies), orbital eccentricity ($e$), longitude of the ascending node ($\Omega$, $s_i$ frequencies), and the obliquity of the axial tilt ($\varepsilon$). For $\varpi$ and $\Omega$, we express our results in terms of $k = e \cos \varpi$ and $q = \sin (i/2) \cos \Omega$, where $i$ represents the orbital inclination, to detect the eigenfrequencies of the apsidal and nodal precession.


\subsection{Baseline Dynamics}
\label{baseline}

\begin{figure*}
  \begin{center}
    \includegraphics[width=\textwidth]{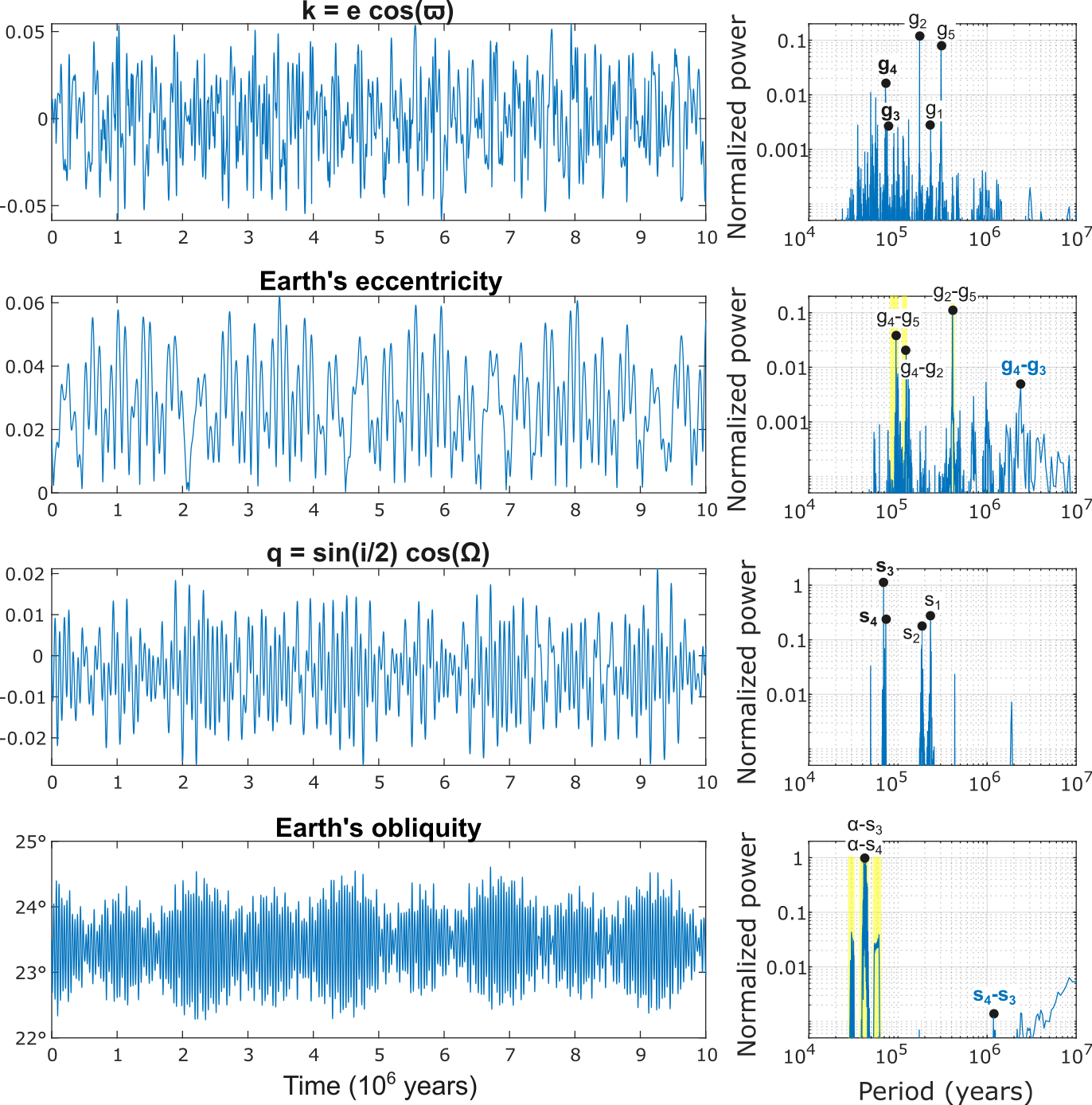}
  \end{center}
  \caption{Earth orbital evolution results, where the mass of Mars is set to 100\% of its present value. The panels show, from top to bottom, the longitude of perihelion in terms of the $k$ parameter, the orbital eccentricity, the longitude of the ascending node in terms of the $q$ parameter, and the obliquity of the axial tilt. The panels on the right shown the normalized Fourier power, with major peaks identified in terms of the $g_i$ and $s_i$ nodes.}
  \label{fig:baseline}
\end{figure*}

To test the reliability of our dynamical simulations, we examined the results for the present Solar System scenario, referred to as the "baseline" scenario, where the mass of Mars is set to 100\% of its current value. Shown in the four panels of Figure~\ref{fig:baseline} are the first $10^7$~years of the $10^8$~years total integration time for (from top to bottom panel) $k$, $e$, $q$, and $\varepsilon$, respectively, with numerous periodic signals visible on this timescale. The panels on the right show the normalized Fourier power, where the associated $g_i$ and $s_i$ nodes have been identified. The Fourier analyses reveal that all four quantities exhibit the expected periodicities and envelopes known from astronomical solutions \citep[e.g.,][]{laskar2004c,laskar2011a} and from stratigraphic detection of orbital forcing \citep[e.g.,][]{hays1976,berger1978c,laskar2004c,laskar2011a,hinnov2018,olsen2019,mogavero2021}. Below we summarize the principal identifications from the baseline simulations.

Longitude of perihelion ($\varpi$): The spectrum of $k = e \cos \varpi$ shows strong power near $P\simeq174$~kyr (consistent with $g_2$) and $\simeq304$~kyr (consistent with $g_5$), with a tight cluster at $P\simeq73$~kyr (consistent with $g_4$ at 72~kyr and $g_3$ at 75~kyr). These features are similar to the modern astronomical solutions \citep{murray1999a,laskar2004c}. A weaker component in the 390--420~kyr range reflects coupling to the $g_2-g_5$ long-eccentricity family.

Eccentricity ($e$): We recover three primary eccentricity signals. 1. A high-power signal at a period of $P\simeq410$~kyr, consistent with the long-eccentricity ``metronome'' associated with $g_2-g_5$ (nominally $405$~kyr) \citep{laskar2004c,olsen2019}. 2. Two short-eccentricity clusters at $\sim$94--99~kyr and $\sim$123--130~kyr, matching the familiar 100-kyr band structure that arises from interference among inner-planet $g$-modes \citep{laskar2004c}. 3. Low-frequency modulation near $\sim$0.971~Myr and $\sim$2.381~Myr, consistent with the Myr-scale Earth--Mars secular beats (the $g_4-g_3$ family) that modulate eccentricity amplitude and phase \citep{laskar2004c,olsen2019}. We note that small offsets from the nominal $405$~kyr value are to be expected given finite-window resolution and long-period amplitude modulation.

Longitude of the ascending node ($\Omega$): As expected for nodal precession, the tallest peak sits at $P\simeq69$~kyr (consistent with $s_3$), with additional lines at $P\simeq73$~kyr (consistent with $s_4$) and combination/sideband structure around $P\simeq185$~kyr (consistent with $s_2$) and $P\simeq231$~kyr (consistent with $s_1$). This pattern is the standard nodal multiplet reported in the La2004/La2010 solutions \citep{laskar2004c,laskar2011a}.

Obliquity ($\varepsilon$): The obliquity spectrum is dominated by a dense multiplet centered on the lunisolar obliquity period, $P\simeq$40--45~kyr. This is the canonical $\sim$41~kyr tilt cycle stabilized by the Moon \citep{laskar1993b}. In line with expectations, the often-cited $\sim$1.2~Myr signal is present but weak; rather, it typically manifests as an envelope that modulates the 41~kyr band (and is most clearly expressed in $e$ or in climatic precession indices) \citep{laskar2004c}. We do see very weak, very long period features ($\sim$9--10~Myr), which are common low-level background or long-window nutation envelopes.

Overall, the four spectra are consistent with the canonical Milankovitch forcing and with inner Solar System secular theory: the long-eccentricity metronome near 405~kyr, the split 100-kyr band in $e$, the nodal band near 70~kyr in $\Omega$, the $\sim$41~kyr obliquity band in $\varepsilon$, and Myr-scale modulation linked to Earth--Mars coupling in $e$ \citep{berger1978c,laskar2004c,olsen2019}. Taken together, the results validate that the integrations with present-day Martian mass produce the standard Earth spectra used in paleoclimate and astrochronology \citep{hays1976,laskar2011a}.


\subsection{Longitude of Perihelion}
\label{somega}

\begin{figure*}
  \begin{center}
    \includegraphics[width=\textwidth]{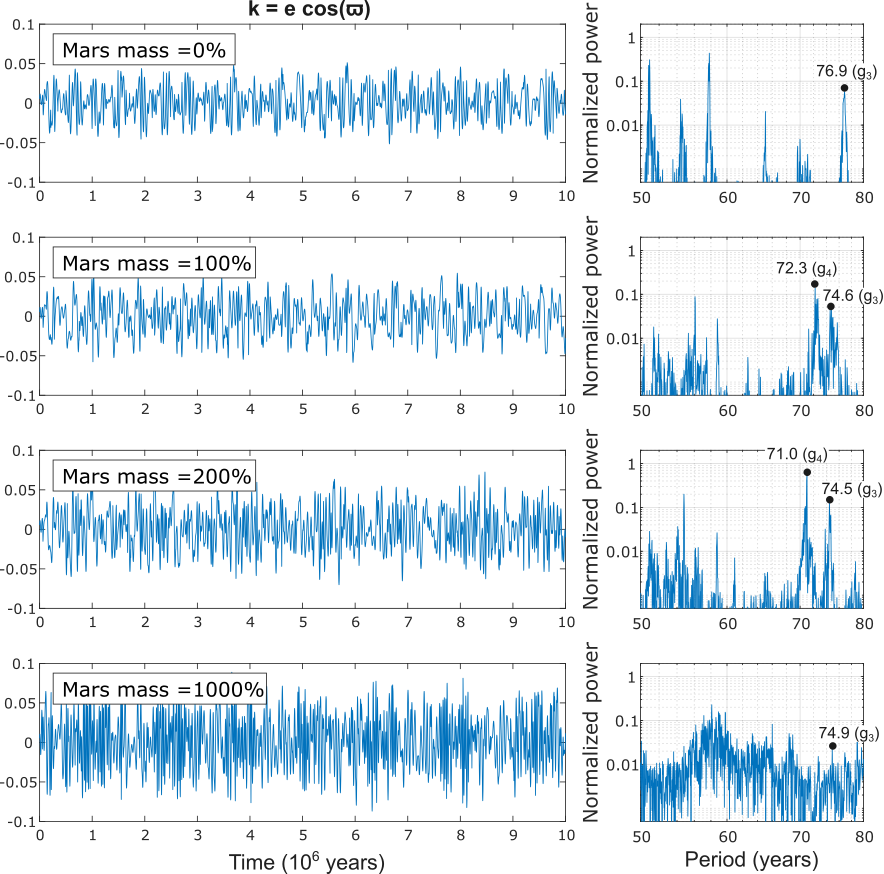}
  \end{center}
  \caption{Periodograms resulting from a Fourier analysis of the $10^8$~yr longitude of perihelion data for Earth, shown for the cases of a 0\%, 100\%, 200\%, and 1000\% Mars mass, from top to bottom, respectively. The panels on the right shown the normalized Fourier power, with major peaks identified in terms of the $g_i$ nodes.}
  \label{fig:sw}
\end{figure*}

For the first dynamical component, the longitude of perihelion, we assessed the data through grouping prominent spectral peaks into canonical bands that identify $g_i$ nodes. Shown in Figure~\ref{fig:sw} are the resulting data and periodograms from the Fourier analysis of the longitude of perihelion data (in terms of $k = e \cos \varpi$) for the specific Mars mass cases of 0\%, 100\%, 200\%, and 1000\%. The detected lines were organized into canonical apsidal/combination bands: the detected lines near $\sim$70~kyr that represent the $g_3$ and $g_4$ frequencies, an apsidal band near $\sim$112~kyr, a prominent combination near $\sim$171~kyr, and a Jupiter–forced apsidal component near $\sim$295~kyr. We also checked for power near $\sim$405~kyr and in Myr-scale windows ($\sim$0.9~Myr and $\sim$2~Myr) to assess possible long-period envelopes known from the eccentricity spectrum and geological orrery constraints.

For the case of no Mars (0\% mass), the $\sim$72~kyr band ($g_4$) is absent entirely while the $g_3$ frequency (of $\sim$75~kyr in the 100\% Mars system) has shifted to 77~kyr. For the case of current Mars (100\% mass), the $\sim$72~kyr band associated with Mars' longitude of perihelion emerges clearly (centroid near $\sim$72.5~kyr; strongest line $\sim$72.3~kyr) along with a slightly weaker $\sim$74.8~kyr band that is linked to Earth's longitude of perihelion. These features are described in more detail in Section~\ref{baseline}. Doubling the mass of Mars (200\%) further strengthens the power of the $g_4$ band, whilst the period is reduced to $\sim$71~kyr. The power of the $\sim$75~kyr band also increases as the mass of Mars increases, resulting in a clearer and sharper $g_3$ peak with a period of $\sim$74.5~kyr. For the 1000\% mass case, the $g_3$ and $g_4$ bands are fragmented into many weak lines. Poorly defined peaks are present around $\sim$68~kyr and $\sim$75~kyr, but pinpointing an exact periodicity associated with $g_3$ and $g_4$ is challenging.

Three robust patterns emerge across the range of Mars mass scenarios. First, the $\sim$70~kyr $g_3$ and $g_4$ bands appear with Mars present, strengthening in power up to the $\sim$200\% Mars mass scenario, and the periods are systematically shortening from $\sim$72~kyr at sub-Mars masses toward $\sim$68~kyr at extreme mass. This reflects the increasing contribution of the Martian $g_4$ and $s_4$ modes to Earth’s apsidal–nodal coupling as Mars’ mass grows. Second, the $\sim$171~kyr band shifts to shorter period and weakens as Mars’ mass increases, ultimately vanishing from the 160--185~kyr window at 1000\% mass. This indicates a re-partitioning of Earth’s apsidal power as the proximity and coupling of relevant $g$-modes are re-tuned by the larger gravitaional influence of Mars. Third, the Jupiter–anchored $\sim$295~kyr band remains near-constant in period (as expected from the stability of $g_5$) but its amplitude declines at high Mars mass, consistent with redistribution of power into nearby combinations when inner–planet couplings are strong. The $\sim$112~kyr band shows no monotonic period trend but a broad decline in summed power with increasing mass, suggesting that part of its content is siphoned into the strengthening $\sim$70~kyr and other combination bands as $g_4$ moves.

Unlike the eccentricity spectrum, where Myr-scale Earth–Mars beats ($g_{4}-g_{3}$) appear as conspicuous low-frequency power, the $\varpi$ spectra show no comparably strong, narrow lines near $\sim$2.4~Myr. This is an expected result since the grand cycles primarily modulate amplitudes and phases of apsidal terms rather than producing sharp standalone lines in $\varpi$ \citep{laskar2004c,laskar2011a,hinnov2018}. Power near $\sim$405~kyr is likewise weak or absent in $\varpi$, since the 405~kyr “metronome” is fundamentally an eccentricity feature tied to $g_{2}-g_{5}$, though minor leakage can occur depending on windowing and sideband structure.


\subsection{Eccentricity}
\label{ecc}

\begin{figure*}
  \begin{center}
    \includegraphics[width=\textwidth]{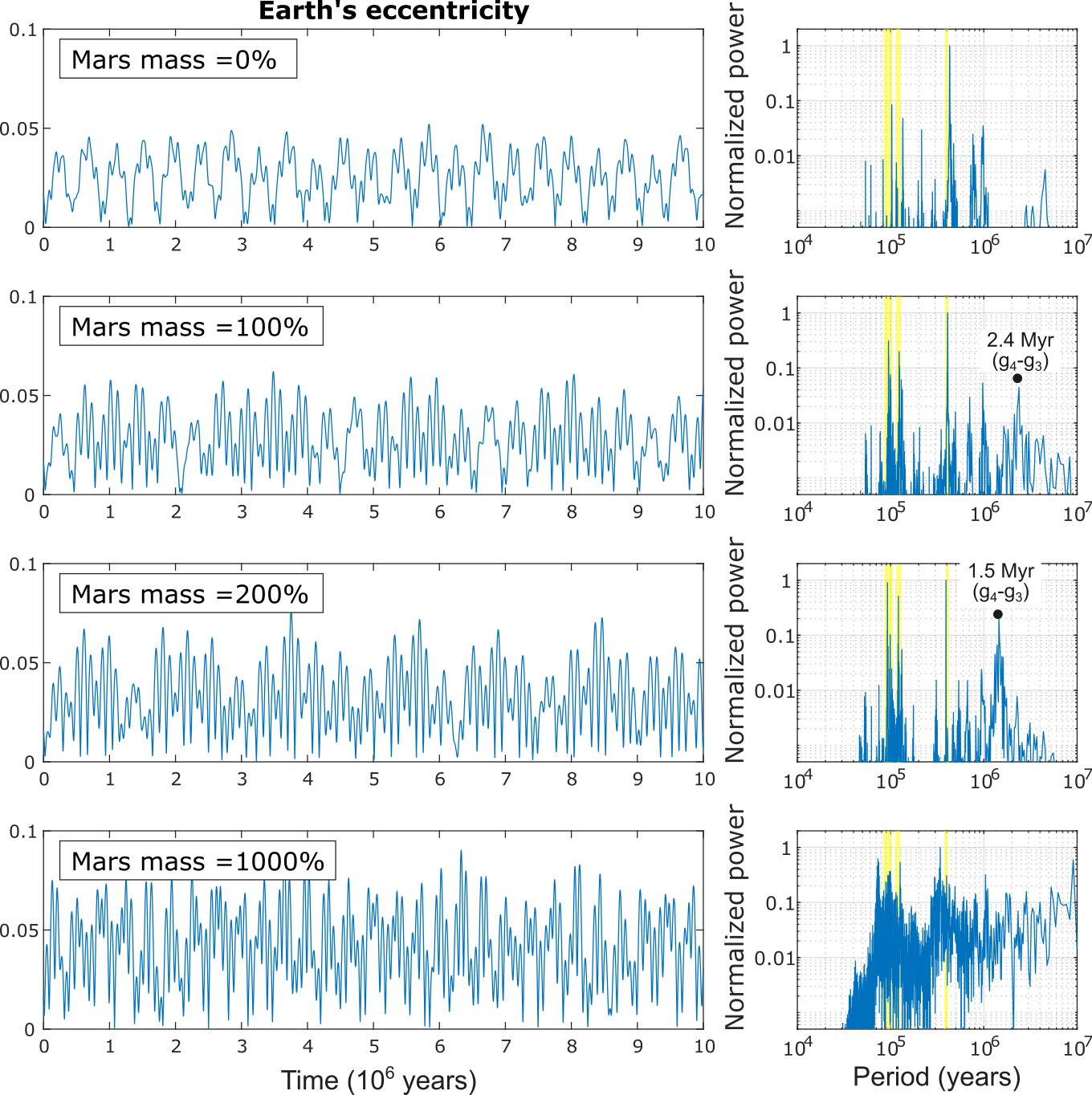}
  \end{center}
  \caption{Periodograms resulting from a Fourier analysis of the $10^8$~yr eccentricity data for Earth, shown for the cases of a 0\%, 100\%, 200\%, and 1000\% Mars mass, from top to bottom, respectively. The panels on the right show the normalized Fourier power, with major peaks identified in terms of the $g_i$ nodes.}
  \label{fig:ecc}
\end{figure*}

Second, we consider the variations in eccentricity that occur across varying Mars' mass. Shown in Figure~\ref{fig:ecc} are the resulting data and periodograms from the Fourier analysis of the eccentricity data for the specific Mars mass cases of 0\%, 100\%, 200\%, and 1000\%. To examine the full suite of periodogram data, we organized the detected lines into canonical bands: (i) the long–eccentricity ``metronome'' near $\sim$405~kyr (the $g_2-g_5$ Venus–Jupiter beat), (ii) the split short–eccentricity band near $\sim$95--100~kyr and $\sim$120--130~kyr (interference among inner-planet $g$-modes), and (iii) Myr-scale envelopes at $\sim$0.9~Myr (the $g_1-g_5$ family) and in the $\sim$1.6--2.8~Myr range associated with Earth–Mars secular beating (the $g_4-g_3$ family).

For the case of no Mars (0\% mass), the spectrum lacks power in the $\sim$1.6--2.8~Myr window, resulting in the loss of the Earth–Mars $g_4-g_3$ grand cycle. The short–eccentricity bands shift from $\sim$95~kyr ($g_4$-$g_5$) and $\sim$123~kyr ($g_4$-$g_2$) to longer periods that center near $\sim$101--102~kyr ($g_3$-$g_5$) and $\sim$133--134~kyr ($g_3$-$g_2$), with reduced total power relative to the other simulations that contain Mars. The 405~kyr family ($g_2$-$g_5$) remains strong and comparatively sharp. For the case of current Mars (100\% mass), the patterns described in Section~\ref{baseline} emerge: a prominent Myr-scale modulation (peaks near $\sim$0.9~Myr and $\sim$2.4~Myr), a well-resolved split of the short–eccentricity band (clusters at $\sim$95--100 and $\sim$120--130~kyr), and a high-power long–eccentricity line whose apparent centroid lies slightly left of 405~kyr due to finite-window leakage and slow amplitude modulation \citep{laskar2004c}. Doubling the mass of Mars (200\%) strengthens the Myr-scale power and shifts the grand-cycle period as expected from secular theory (the Martian proper frequency $g_4$ moves with perturber mass) from $\sim$2.4-Myr to $\sim$1.4~Myr, while the short–eccentricity bands become a few kyr shorter and gain relative power. For the 1000\% Mars mass case (where Mars is slightly more massive than Earth), low-frequency cycles dominates the Myr domain but do not exhibit a clearly distinguishable periodicity. The short–eccentricity bands become shorter and stronger compared to the 100\% and 200\% runs, whereas the 405~kyr family persists but sits amid richer side-structure arising from intense modulation with its peak shifted to $\sim$340~kyr.

Across the full scan of Mars mass scenarios, three patterns stand out. First, the amplitude of the $g_4-g_3$ family increases roughly monotonically as Mars’ mass increases from 0 to 200\%, and its period shifts systematically (also see Figure~\ref{fig:power}), reflecting the mass dependence of the secular eigenfrequency $g_4$ and its distance from Earth’s $g_3$ \citep{murray1999a,laskar2004c}. Second, both short–eccentricity clusters exhibit a tendency to shorten by several kyr as Mars’ mass increases from 0\% through super-Mars regimes, with a concurrent rise in their summed power. This behavior is consistent with stronger Earth–Mars coupling re-tuning the interference among inner-planet $g$-modes. Third, the 405~kyr metronome is present at all masses, as expected for the robust $g_2-g_5$ difference, while its apparent centroid and local side-structure vary at the percent level, plausibly due to finite spectral resolution, slow envelopes, and leakage when Myr-scale modulation is strong \citep{laskar2004c,hinnov2018}.


\subsection{Longitude of the Ascending Node}
\label{bomega}

\begin{figure*}
  \begin{center}
    \includegraphics[width=\textwidth]{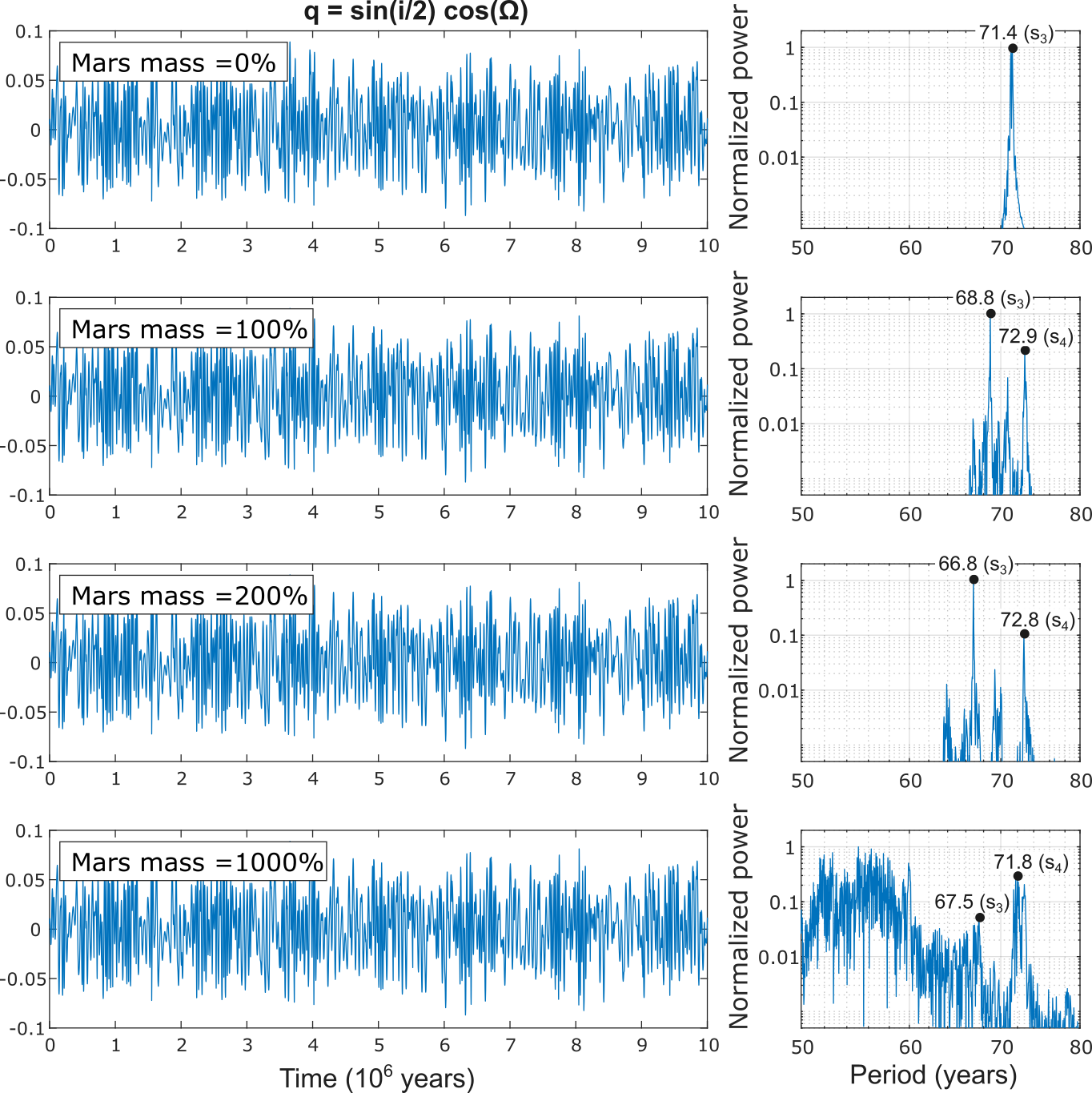}
  \end{center}
  \caption{Periodograms resulting from a Fourier analysis of the $10^8$~yr longitude of the ascending node data for Earth, shown for the cases of a 0\%, 100\%, 200\%, and 1000\% Mars mass, from top to bottom, respectively. The panels on the right shown the normalized Fourier power, with major peaks identified in terms of the $s_i$ nodes.}
  \label{fig:bw}
\end{figure*}

The third dynamical component examined was the longitude of the ascending node. Shown in Figure~\ref{fig:bw} are the resulting data and periodograms of the longitude of the ascending node (in terms of $q = \sin (i/2) \cos \Omega$) for the specific Mars mass cases of 0\%, 100\%, 200\%, and 1000\%. The detected lines in the power spectra were again organized into the dynamical families expected from linear secular theory, as described in Section~\ref{baseline}. The dynamical family groupings consisted of the dominant nodal band (Earth's $s$-mode content) in the $\sim$45--80~kyr window, and two combination clusters near $\sim$170~kyr and $\sim$220--230~kyr that may result from mixing nodal and apsidal frequencies of the terrestrial planets. We also inspected longer periods for weak coupling to the long-eccentricity family (near 405~kyr) and for Myr-scale envelopes known to modulate other orbital elements.

For the case of no Mars (0\% mass), the nodal band is narrow and strong at $\sim$71~kyr ($s_3$), the $\sim$170~kyr and $\sim$220--230~kyr combination clusters are present, and there is negligible power near 405~kyr or at Myr periods. For the case of present Mars (100\% mass), the $s_3$ band remains strong but shifts to slightly shorter periods reaching $\sim$68.8~kyr. A second strong peak is linked to $s_4$ at $\sim$72.9~kyr that is absent in the 0\% case. Doubling Mars' mass (200\%) accelerates nodal precession further (shorter periods in the dominant band) and shifts the $s_3$ and $s_4$ bands to even shorter periods, $\sim$66.8~kyr and $\sim$72.8~kyr, respectively. For the case of an Earth-mass Mars (1000\% mass), the nodal band shortens in period markedly and fragments into many weaker components. A clear $s_4$ peak is detected around $\sim$72~kyr, but the $s_3$ peak is not clearly defined, estimated to lie around $\sim$67.5~kyr. At this extreme mass, we also see weak, broad Myr-scale features, consistent with slow envelopes rather than sharp precessional lines.

Three systematic patterns emerge across the Mars mass simulation results. First, the period centroid of the dominant nodal band monotonically shortens as Mars’ mass increases: from near the canonical $\sim$72.9~kyr at 100\% mass toward shorter values by 200\%, and still shorter to $\sim$71.8~kyr in the 1000\% case. This reflects the increasing contribution of the Martian proper mode $s_4$ and its coupling into Earth’s nodal motion. Second, the $\sim$170~kyr and $\sim$220--230~kyr clusters migrate to shorter periods and change amplitude as the mass of Mars increases, indicating a re-partitioning of apsidal-nodal combination power as the proximity of the relevant $g$ and $s$-modes are re-tuned. Third, power near $\sim$405~kyr remains weak or absent in $\Omega$ at all masses, consistent with the fact that the 405~kyr “metronome’’ is fundamentally an eccentricity feature tied to $g_{2}-g_{5}$.


\subsection{Obliquity}
\label{obl}

\begin{figure*}
  \begin{center}
    \includegraphics[width=\textwidth]{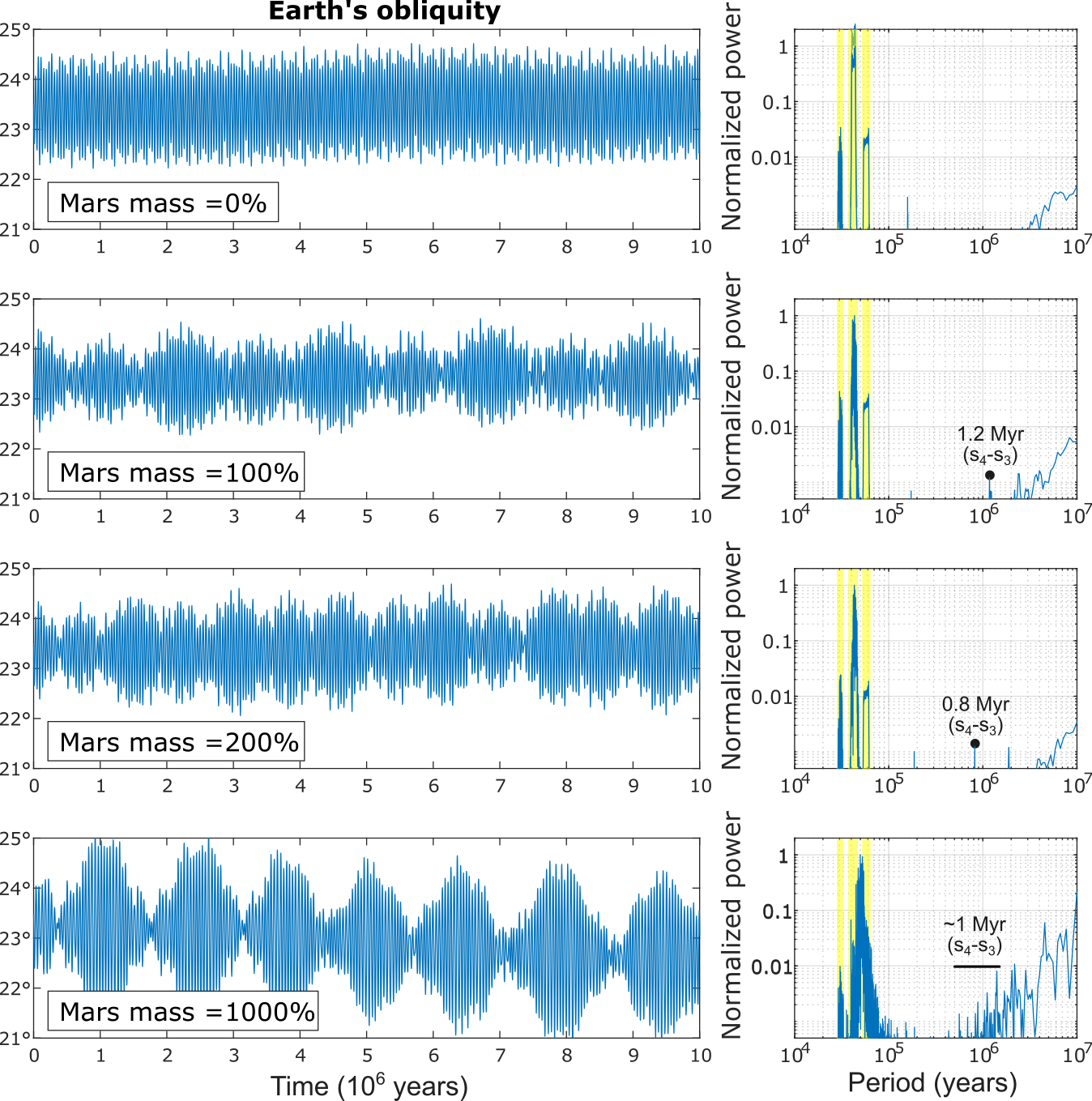}
  \end{center}
  \caption{Periodograms resulting from a Fourier analysis of the $10^8$~yr obliquity data for Earth, shown for the cases of a 0\%, 100\%, 200\%, and 1000\% Mars mass, from top to bottom, respectively. The panels on the right shown the normalized Fourier power, with major peaks identified in terms of the $s_i$ nodes.}
  \label{fig:obl}
\end{figure*}

Finally, we examine the data tracking the variations in Earth's obliquity cycles as a function of Mars mass. Shown in Figure~\ref{fig:obl} are the resulting data and periodograms of Earth's obliquity for the specific Mars mass cases of 0\%, 100\%, 200\%, and 1000\%. The period detections were organized into the following dynamical bands: the principal lunisolar obliquity band near $\sim$41~kyr; neighboring sidebands on either side (low: 29–-32~kyr; high: 54–-60~kyr); and low-frequency envelopes near $\sim$0.9~Myr, $\sim$1.2~Myr, and $\sim$10~Myr that typically manifest as amplitude modulations of the 41~kyr band rather than as sharp, standalone lines in $\varepsilon$ \citep{laskar1993b}.

For the no Mars case (0\% mass), the spectrum is dominated by the canonical obliquity band; a power-weighted centroid near 42~kyr, with the strongest peak at $\sim$43.7~kyr, and with substantial integrated power confined to 39-–44~kyr. This results from the interaction between the $s_3$ cycle and the rotation of Earth's axial precession ($\alpha$). A very long period feature near $\sim$9–-10~Myr is present, but has modest integrated power, consistent with a weak envelope. At 100\% Mars mass, the 42~kyr peak remains the most powerful, but we also see evidence of a $s3-s4$ beat period near $\sim$1.2~Myr. Doubling the mass of Mars (200\%) results in a similar period of the main cycle, with a centroid located at 43.2~kyr. The $s3-s4$ beat period has shifted to $\sim$0.8~Myr. In the 1000\% Mars mass case, the canonical 39–-44~kyr band shifts to $\sim$39–-65~kyr, with the strongest peak near $\sim$50~kyr. Very long ($\sim$10~Myr) peaks become prominent at this extreme mass, signaling strong slow envelopes but there is no one clear beat period peak that can be identified among the many smaller fragmented frequencies.

There are several obliquity patterns that emerge from an analysis of the full range of Mars mass simulations. First, the characteristic obliquity period lengthens with increasing Mars mass, and then power moves out of the canonical window altogether at 1000\% mass, producing a dominant cycle near $\sim$50~kyr. Second, the high sideband (54–-60~kyr) intensifies from 0\% to 200\% and merges into a broad 39–-65~kyr complex that ultimately dominates at a Mars mass of 1000\%. Third, power at the $\sim$0.9–-1.2~Myr window remains weak as sharp lines at all masses, consistent with the understanding that these timescales act primarily as amplitude envelopes of the obliquity band, except at extreme masses where very long ($\sim$10~Myr) periodicity becomes prominent.


\section{Discussion}
\label{discussion}

Studying Solar System style orbital dynamics in a controlled way offers a uniquely high-fidelity laboratory for exoplanet science. It allows us to take advantage of the one planetary system we know in intimate detail to build our understanding of various processes that can shape a particular planetary system, and which might prove pivotal in future efforts to search for evidence of life beyond the Solar System \citep[e.g.,][and references therein]{horner2010e,horner2020b}. In particular, varying the mass and placement of a single perturber (here, Mars) cleanly exposes how secular modes, resonances, and spin–orbit coupling sculpt the long-term evolution of terrestrial worlds; insights that transfer directly to the interpretation of compact multi-planet systems and terrestrial planets residing alongside giant companions \citep[e.g.,][]{horner2020a,horner2020b,kane2023a}. At the population level, the architectures of Kepler multi-planet systems emphasize the prevalence of tightly packed yet quasi-stable configurations in or near resonance, where weak secular coupling accumulates over Myr–-Gyr timescales to shape obliquity and eccentricity driven climates \citep[e.g.,][]{menou2003a,lissauer2011b,fabrycky2014}. These threads motivate using Solar System analog experiments as a bridge between detailed celestial mechanics and exoplanet habitability assessment.

\begin{figure*}
  \begin{center}
    \begin{tabular}{cc}   
      \includegraphics[width=\columnwidth]{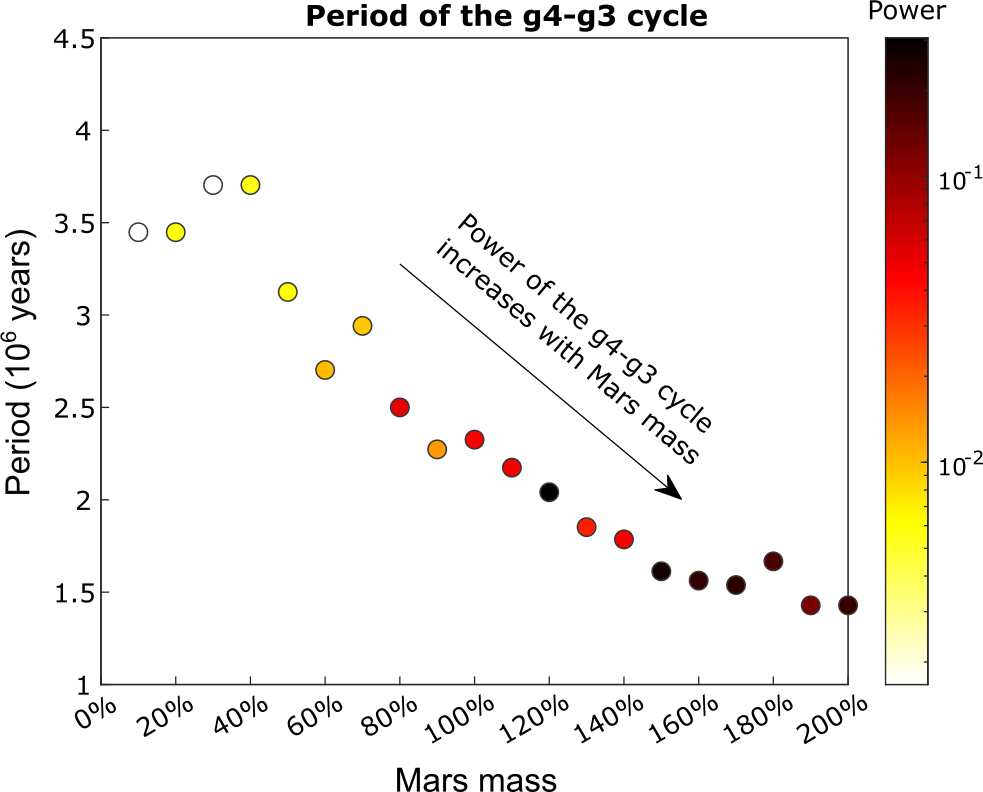} &
      \includegraphics[width=\columnwidth]{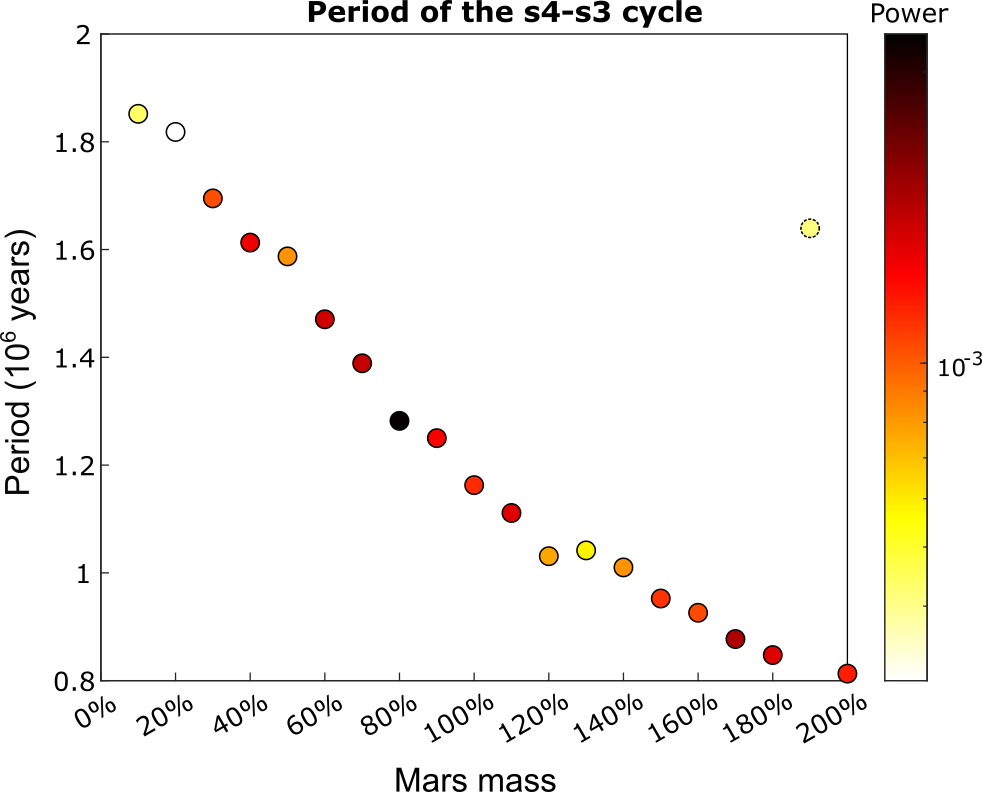}
    \end{tabular}
  \end{center}
  \caption{Left: The period and relative Fourier power (normalized to the strongest 400~kyr eccentricity cycle) of the $g_4-g_3$ cycle as a function of the mass of Mars. Right: The period and relative Fourier power (normalized to the strongest 41~kyr primary obliquity cycle) of the $s_4-s_3$ cycle as a function of the mass of Mars.}
  \label{fig:power}
\end{figure*}

Shown in Figure~\ref{fig:power} are the periods and normalized Fourier powers of the $g_4-g_3$  and $s_4-s_3$ cycles for each of the Mars masses explored in this work. Overall, our results show that the period of the $g_4-g_3$ cycle decreases whilst the relative Fourier power increases with increasing Mars mass. Although the period of the $s_4-s_3$ cycle also decreases with increasing Mars mass, the strength of the Fourier power varies significantly. For the case of 0\% Mars mass, the $g_4-g_3$ and $s_4-s_3$ beat periods in eccentricity and obliquity disappear as expected. For the case of 100\% Mars mass, the 2.4~Myr eccentricity cycle and 1.2~Myr obliquity cycle are in 2:1 resonance, which serves to amplify those beat periods. For the case of 200\% Mars mass, the 1.5~Myr eccentricity cycle and 0.8~Myr obliquity cycle are not quite in 2:1 resonance so the beat periods are not as pronounced. When the Mars mass is raised to 1000\%, the $g$ and $s$ frequencies are less well defined, so identifying the presence of resonance between $g_4-g_3$ and $s_4-s_3$ is challenging. However, a significant beat period present in the obliquity time series along with several stand-alone peaks in its periodogram near $\sim$1~Myr suggests alternating $s_4-s_3$ beat periods (see also Figure~\ref{fig:obl}).

The results of our simulations are largely consistent with the predictions of linear secular theory and its modern extensions, since the proper apsidal frequencies $g_i$ scale with planetary masses and semi-major axis ratios \citep{murray1999a}. Consequently, altering Mars’ mass predominantly shifts $g_4$ and enhances coupling into Earth’s $g_3$ component. For Earth's eccentricity, the observed growth and period drift of the Myr-scale Earth–Mars beat with increasing Martian mass is a direct result of this enhanced coupling. Because climatic precession forcing scales with eccentricity \citep{dressing2010}, these reorganizations of the short–eccentricity band and the strengthening of Myr-scale envelopes imply correspondingly stronger influences of climatic precession cycles on the long-term climate evolution with a more massive Mars. Conversely, and as noted above, in the limit of vanishing Mars, the eccentricity spectrum simplifies, resulting in a loss of the Earth–Mars grand cycle, the weakening and lengthening of the short–eccentricity bands, and causing the long–eccentricity metronome to stand out.

In the case of Earth’s obliquity, the response to changing the mass of Mars is governed by the tuning between the spin precession constant, $\alpha$ (set chiefly by lunisolar torques and Earth’s dynamical form-factor $J_2$), and the relevant orbital nodal eigenfrequencies $s_i$. The forcing enters through combinations of the form $(\alpha - s_i)$, with coupling strengths determined by the Laplace–Lagrange coefficients \citep{murray1999a}. Shifting the Martian proper modes $(g_4,s_4)$ and strengthening their coupling into the terrestrial mode re-tunes $(\alpha - s)$ that sets the obliquity band. The observed 41~kyr cycle at moderate Mars mass differs from the $\sim$50~kyr cycles at extreme Mars mass, following directly from this re-tuning mechanism. The rise of very long envelopes at 1000\% mass is a result of strong coupling and near-commensurabilities that modulate the tilt-band amplitude without necessarily introducing new sharp obliquity lines \citep{laskar1993b,laskar2004c,hinnov2018}. The narrow but weak $\sim$1.2~Myr features in $\varepsilon$ across the Mars mass scan therefore reflects the general result that Myr-scale ``grand cycles’’ are most visible as envelopes rather than as discrete obliquity frequencies \citep{laskar2004c,laskar2011a}.

In exoplanet contexts, climate and habitability respond sensitively to the statistics of the calculated and inferred Milankovitch cycles. Elevated or time-variable eccentricity can warm terrestrial planets and avert snowball states; high-amplitude obliquity cycles can both redistribute insolation and, in some regimes, suppress ice–albedo feedback, expanding the viable parameter space for surface liquid water \citep[e.g.,][]{williams2002,spiegel2009a,dressing2010,spiegel2010b,kane2012e,armstrong2014b,dobrovolskis2013b,way2017a,kane2017d,kane2021a}. Our simulation suite of varying Mars' mass thus offers an exo-Milankovitch analog: mapping how the periods and amplitudes of a terrestrial planet’s apsidal and nodal precession, eccentricity, and obliquity cycles respond to changes in the system's planet-planet forcing. The specific dependence on how a Mars-type perturber shifts the dominant $g$ and $s$ modes and alters spin precession commensurabilities directly informs which system architectures are most likely to yield clement climates on Earth-like exoplanets. 

Beyond climate, the same couplings bear on long-term orbital stability and observational diagnostics. Secular spin–orbit resonances can trap a planet at large obliquity or drive chaotic obliquity diffusion, with outcomes controlled by the masses and orbits of neighboring giants; obliquity tides and resonance-maintained tilts may even sculpt planetary system architectures and thermal states \citep[e.g.,][]{atobe2004,atobe2007,saillenfest2019a,millholland2019a}. Case studies applying these ideas to specific systems (e.g., Kepler-62f) show that the presence of external giants can transform an ostensibly Earth-similar world's spin–orbit history and climate \citep{williams1997b,quarles2020a,kane2017d,kane2021a}. In cases where giant planets contribute significantly to dynamical forcing, the mapping of dynamical periods and cycles provides insight into the requirements for generalizing Milankovitch theory to exoplanets \citep{deitrick2018a,deitrick2018b,vervoort2022}. However, it is worth noting that radial velocity surveys have revealed that giant planets beyond the snow line are relatively rare, even for solar-type stars \citep[e.g.,][]{wittenmyer2011a,wittenmyer2016c,wittenmyer2020b,fulton2021,rosenthal2021,bonomo2023}, emphasizing the importance of properly accounting for giant-planet forcing dynamical components. In this regard, our Mars mass experiments supply a calibrated template for how increasing the mass of a modestly exterior perturber shortens or lengthens key precession periods, shifts resonant crossings, and changes obliquity forcing spectra. Such templates aid target triage (which systems likely host climate-stable terrestrials), guide interpretation of observed eccentricities/obliquities, and help prioritize follow-up for worlds whose Milankovitch statistics predict favorable, persistent habitability. 


\section{Conclusions}
\label{conclusions}

As the scientific discourse regarding planetary habitability factors continue, the contribution of dynamical components to terrestrial planet climate evolution, and their detectability for exoplanets, have taken on an increasing importance in the discussion. Our dynamical simulations map how Earth's principal Milankovitch bands respond to a controlled change in the mass of Mars, using long integrations of Earth's longitude of perihelion ($\varpi$), eccentricity ($e$), longitude of the ascending node ($\Omega$), and obliquity ($\varepsilon$). By scanning Mars' mass from 0\% to 1000\% of its present value, we isolated how the inner Solar System secular architecture re-tunes as a function of Mars mass. Overall, the observed trends for the four dynamical components are those anticipated for a system in which a tunable Mars-type perturber reorganizes the inner Solar System secular architecture. The results provide a calibrated ``response function" that connects planetary architecture to the periods and amplitudes of the cycles that pace climate. Although our work only considers the specific effects of alternate Mars masses, opportunities for further investigations include varying the semi-major axis and inclination of Mars' orbit, the use of climate models with the measured forcing histories to evaluate ice–albedo feedback, snowball avoidance, and habitability persistence across architecture families, and applying the resulting templates to specific exoplanet targets. Additionally, the presented simulations would ideally solve the obliquity calculations simultaneously with the N-body calculations, but this would not have a significant effect on our results.

Our simulations have revealed several important points. Some pacing agents, notably the $\sim\!405$~kyr cycle driven by Jupiter's apsidal precession, are architecturally robust for the range of Mars masses explored, while the detailed morphology, centroids, and amplitudes of the $\sim$100~kyr bands are highly sensitive to the mass (and hence the $g_i,s_i$ eigenfrequencies and coupling) of a nearby terrestrial perturber. Climatic precession forcing (via $e$ and $\varpi$) and obliquity forcing therefore depend in a predictable way on system architecture. For the particular scenarios described here, increasing the mass of an exterior Mars-like planet strengthens Myr-scale modulation, lengthens short–eccentricity and obliquity bands, and re-partitions power among apsidal–nodal combinations. An important implication is that, despite its small mass, a Mars-type planet exerts a non-negligible amount of influence on the climate of other terrestrial planets. These calibrated trends are directly portable to exoplanetary systems when assessing exo–Milankovitch pacing, orbital stability margins, and prospects for long-lived clement climates. Our findings thus motivate a broader program that links exoplanet architectures to orbital–spin forcing statistics and, ultimately, to the climates and habitability of terrestrial worlds.


\section*{Acknowledgements}

The results reported herein benefited from collaborations and/or information exchange within NASA's Nexus for Exoplanet System Science (NExSS) research coordination network sponsored by NASA's Science Mission Directorate.


\software{Mercury \citep{chambers1999}, La1993 obliquity code \citep{laskar1993b} accessible via http://vizier.cfa.harvard.edu/ftp/cats/VI/63/.}



\begin{thebibliography}{}
\expandafter\ifx\csname natexlab\endcsname\relax\def\natexlab#1{#1}\fi
\providecommand{\url}[1]{\href{#1}{#1}}
\providecommand{\dodoi}[1]{doi:~\href{http://doi.org/#1}{\nolinkurl{#1}}}
\providecommand{\doeprint}[1]{\href{http://ascl.net/#1}{\nolinkurl{http://ascl.net/#1}}}
\providecommand{\doarXiv}[1]{\href{https://arxiv.org/abs/#1}{\nolinkurl{https://arxiv.org/abs/#1}}}

\bibitem[{{Andrews-Hanna} {et~al.}(2008){Andrews-Hanna}, {Zuber}, \&
  {Banerdt}}]{andrewshanna2008a}
{Andrews-Hanna}, J.~C., {Zuber}, M.~T., \& {Banerdt}, W.~B. 2008, \nat, 453,
  1212, \dodoi{10.1038/nature07011}

\bibitem[{{Armstrong} {et~al.}(2014){Armstrong}, {Barnes}, {Domagal-Goldman},
  {Breiner}, {Quinn}, \& {Meadows}}]{armstrong2014b}
{Armstrong}, J.~C., {Barnes}, R., {Domagal-Goldman}, S., {et~al.} 2014,
  Astrobiology, 14, 277, \dodoi{10.1089/ast.2013.1129}

\bibitem[{{Atobe} \& {Ida}(2007)}]{atobe2007}
{Atobe}, K., \& {Ida}, S. 2007, \icarus, 188, 1,
  \dodoi{10.1016/j.icarus.2006.11.022}

\bibitem[{{Atobe} {et~al.}(2004){Atobe}, {Ida}, \& {Ito}}]{atobe2004}
{Atobe}, K., {Ida}, S., \& {Ito}, T. 2004, \icarus, 168, 223,
  \dodoi{10.1016/j.icarus.2003.11.017}

\bibitem[{{Berger}(1978)}]{berger1978c}
{Berger}, A.~L. 1978, Journal of the Atmospheric Sciences, 35, 2362,
  \dodoi{10.1175/1520-0469(1978)035<2362:LTVODI>2.0.CO;2}

\bibitem[{{Bitsch} {et~al.}(2015){Bitsch}, {Lambrechts}, \&
  {Johansen}}]{bitsch2015b}
{Bitsch}, B., {Lambrechts}, M., \& {Johansen}, A. 2015, \aap, 582, A112,
  \dodoi{10.1051/0004-6361/201526463}

\bibitem[{{Bonomo} {et~al.}(2023){Bonomo}, {Dumusque}, {Massa}, {Mortier},
  {Bongiolatti}, {Malavolta}, {Sozzetti}, {Buchhave}, {Damasso}, {Haywood},
  {Morbidelli}, {Latham}, {Molinari}, {Pepe}, {Poretti}, {Udry}, {Affer},
  {Boschin}, {Charbonneau}, {Cosentino}, {Cretignier}, {Ghedina}, {Lega},
  {L{\'o}pez-Morales}, {Margini}, {Mart{\'\i}nez Fiorenzano}, {Mayor},
  {Micela}, {Pedani}, {Pinamonti}, {Rice}, {Sasselov}, {Tronsgaard}, \&
  {Vanderburg}}]{bonomo2023}
{Bonomo}, A.~S., {Dumusque}, X., {Massa}, A., {et~al.} 2023, \aap, 677, A33,
  \dodoi{10.1051/0004-6361/202346211}

\bibitem[{{Brasser} {et~al.}(2009){Brasser}, {Morbidelli}, {Gomes}, {Tsiganis},
  \& {Levison}}]{brasser2009}
{Brasser}, R., {Morbidelli}, A., {Gomes}, R., {Tsiganis}, K., \& {Levison},
  H.~F. 2009, \aap, 507, 1053, \dodoi{10.1051/0004-6361/200912878}

\bibitem[{{Chambers}(1999)}]{chambers1999}
{Chambers}, J.~E. 1999, \mnras, 304, 793,
  \dodoi{10.1046/j.1365-8711.1999.02379.x}

\bibitem[{{Clement} {et~al.}(2019){Clement}, {Kaib}, {Raymond}, {Chambers}, \&
  {Walsh}}]{clement2019b}
{Clement}, M.~S., {Kaib}, N.~A., {Raymond}, S.~N., {Chambers}, J.~E., \&
  {Walsh}, K.~J. 2019, \icarus, 321, 778, \dodoi{10.1016/j.icarus.2018.12.033}

\bibitem[{{Clement} {et~al.}(2018){Clement}, {Kaib}, {Raymond}, \&
  {Walsh}}]{clement2018}
{Clement}, M.~S., {Kaib}, N.~A., {Raymond}, S.~N., \& {Walsh}, K.~J. 2018,
  \icarus, 311, 340, \dodoi{10.1016/j.icarus.2018.04.008}

\bibitem[{{Dauphas} \& {Pourmand}(2011)}]{dauphas2011b}
{Dauphas}, N., \& {Pourmand}, A. 2011, \nat, 473, 489,
  \dodoi{10.1038/nature10077}

\bibitem[{{Deitrick} {et~al.}(2018{\natexlab{a}}){Deitrick}, {Barnes}, {Quinn},
  {Armstrong}, {Charnay}, \& {Wilhelm}}]{deitrick2018a}
{Deitrick}, R., {Barnes}, R., {Quinn}, T.~R., {et~al.} 2018{\natexlab{a}}, \aj,
  155, 60, \dodoi{10.3847/1538-3881/aaa301}

\bibitem[{{Deitrick} {et~al.}(2018{\natexlab{b}}){Deitrick}, {Barnes}, {Bitz},
  {Fleming}, {Charnay}, {Meadows}, {Wilhelm}, {Armstrong}, \&
  {Quinn}}]{deitrick2018b}
{Deitrick}, R., {Barnes}, R., {Bitz}, C., {et~al.} 2018{\natexlab{b}}, \aj,
  155, 266, \dodoi{10.3847/1538-3881/aac214}

\bibitem[{{Dobrovolskis}(2013)}]{dobrovolskis2013b}
{Dobrovolskis}, A.~R. 2013, \icarus, 226, 760,
  \dodoi{10.1016/j.icarus.2013.06.026}

\bibitem[{{Dressing} {et~al.}(2010){Dressing}, {Spiegel}, {Scharf}, {Menou}, \&
  {Raymond}}]{dressing2010}
{Dressing}, C.~D., {Spiegel}, D.~S., {Scharf}, C.~A., {Menou}, K., \&
  {Raymond}, S.~N. 2010, \apj, 721, 1295, \dodoi{10.1088/0004-637X/721/2/1295}

\bibitem[{{Dutkiewicz} {et~al.}(2024){Dutkiewicz}, {Boulila}, \& {Dietmar
  M{\"u}ller}}]{dutkiewicz2024}
{Dutkiewicz}, A., {Boulila}, S., \& {Dietmar M{\"u}ller}, R. 2024, Nature
  Communications, 15, 1998, \dodoi{10.1038/s41467-024-46171-5}

\bibitem[{{Edvardsson} {et~al.}(2002){Edvardsson}, {Karlsson}, \&
  {Engholm}}]{edvardsson2002}
{Edvardsson}, S., {Karlsson}, K.~G., \& {Engholm}, M. 2002, \aap, 384, 689,
  \dodoi{10.1051/0004-6361:20020029}

\bibitem[{{Fabrycky} {et~al.}(2014){Fabrycky}, {Lissauer}, {Ragozzine}, {Rowe},
  {Steffen}, {Agol}, {Barclay}, {Batalha}, {Borucki}, {Ciardi}, {Ford},
  {Gautier}, {Geary}, {Holman}, {Jenkins}, {Li}, {Morehead}, {Morris},
  {Shporer}, {Smith}, {Still}, \& {Van Cleve}}]{fabrycky2014}
{Fabrycky}, D.~C., {Lissauer}, J.~J., {Ragozzine}, D., {et~al.} 2014, \apj,
  790, 146, \dodoi{10.1088/0004-637X/790/2/146}

\bibitem[{{Ford}(2014)}]{ford2014}
{Ford}, E.~B. 2014, Proceedings of the National Academy of Science, 111, 12616,
  \dodoi{10.1073/pnas.1304219111}

\bibitem[{{Fulton} {et~al.}(2021){Fulton}, {Rosenthal}, {Hirsch}, {Isaacson},
  {Howard}, {Dedrick}, {Sherstyuk}, {Blunt}, {Petigura}, {Knutson}, {Behmard},
  {Chontos}, {Crepp}, {Crossfield}, {Dalba}, {Fischer}, {Henry}, {Kane},
  {Kosiarek}, {Marcy}, {Rubenzahl}, {Weiss}, \& {Wright}}]{fulton2021}
{Fulton}, B.~J., {Rosenthal}, L.~J., {Hirsch}, L.~A., {et~al.} 2021, \apjs,
  255, 14, \dodoi{10.3847/1538-4365/abfcc1}

\bibitem[{{Gilmore} \& {Ross}(2008)}]{gilmore2008}
{Gilmore}, J.~B., \& {Ross}, A. 2008, \prd, 78, 124021,
  \dodoi{10.1103/PhysRevD.78.124021}

\bibitem[{{Hansen}(2009)}]{hansen2009c}
{Hansen}, B. M.~S. 2009, \apj, 703, 1131, \dodoi{10.1088/0004-637X/703/1/1131}

\bibitem[{{Hays} {et~al.}(1976){Hays}, {Imbrie}, \& {Shackleton}}]{hays1976}
{Hays}, J.~D., {Imbrie}, J., \& {Shackleton}, N.~J. 1976, Science, 194, 1121,
  \dodoi{10.1126/science.194.4270.1121}

\bibitem[{{Hinnov}(2018)}]{hinnov2018}
{Hinnov}, L.~A. 2018, Proceedings of the National Academy of Science, 115,
  6104, \dodoi{10.1073/pnas.1807020115}

\bibitem[{{Horner} \& {Jones}(2008)}]{horner2008a}
{Horner}, J., \& {Jones}, B.~W. 2008, International Journal of Astrobiology, 7,
  251, \dodoi{10.1017/S1473550408004187}

\bibitem[{{Horner} \& {Jones}(2010)}]{horner2010e}
---. 2010, International Journal of Astrobiology, 9, 273,
  \dodoi{10.1017/S1473550410000261}

\bibitem[{{Horner} {et~al.}(2020{\natexlab{a}}){Horner}, {Vervoort}, {Kane},
  {Ceja}, {Waltham}, {Gilmore}, \& {Kirtland Turner}}]{horner2020a}
{Horner}, J., {Vervoort}, P., {Kane}, S.~R., {et~al.} 2020{\natexlab{a}}, \aj,
  159, 10, \dodoi{10.3847/1538-3881/ab5365}

\bibitem[{{Horner} {et~al.}(2025){Horner}, {Wittenmyer}, {Kane}, \&
  {Holt}}]{horner2025}
{Horner}, J., {Wittenmyer}, R.~A., {Kane}, S.~R., \& {Holt}, T.~R. 2025, \aj,
  169, 8, \dodoi{10.3847/1538-3881/ad8e3a}

\bibitem[{{Horner} {et~al.}(2020{\natexlab{b}}){Horner}, {Kane}, {Marshall},
  {Dalba}, {Holt}, {Wood}, {Maynard-Casely}, {Wittenmyer}, {Lykawka}, {Hill},
  {Salmeron}, {Bailey}, {L{\"o}hne}, {Agnew}, {Carter}, \&
  {Tylor}}]{horner2020b}
{Horner}, J., {Kane}, S.~R., {Marshall}, J.~P., {et~al.} 2020{\natexlab{b}},
  \pasp, 132, 102001, \dodoi{10.1088/1538-3873/ab8eb9}

\bibitem[{{Ida} {et~al.}(2016){Ida}, {Guillot}, \& {Morbidelli}}]{ida2016a}
{Ida}, S., {Guillot}, T., \& {Morbidelli}, A. 2016, \aap, 591, A72,
  \dodoi{10.1051/0004-6361/201628099}

\bibitem[{{Ikeda} \& {Tada}(2020)}]{ikeda2020a}
{Ikeda}, M., \& {Tada}, R. 2020, Earth and Planetary Science Letters, 537,
  116168, \dodoi{10.1016/j.epsl.2020.116168}

\bibitem[{{Imbrie} {et~al.}(1992){Imbrie}, {Boyle}, {Clemens}, {Duffy},
  {Howard}, {Kukla}, {Kutzbach}, {Martinson}, {McIntyre}, {Mix}, {Molfino},
  {Morley}, {Peterson}, {Pisias}, {Prell}, {Raymo}, {Shackleton}, \&
  {Toggweiler}}]{imbrie1992}
{Imbrie}, J., {Boyle}, E.~A., {Clemens}, S.~C., {et~al.} 1992,
  Paleoceanography, 7, 701, \dodoi{10.1029/92PA02253}

\bibitem[{{Kane}(2023)}]{kane2023a}
{Kane}, S.~R. 2023, \psj, 4, 38, \dodoi{10.3847/PSJ/acbb6b}

\bibitem[{{Kane} \& {Gelino}(2012)}]{kane2012e}
{Kane}, S.~R., \& {Gelino}, D.~M. 2012, Astrobiology, 12, 940,
  \dodoi{10.1089/ast.2011.0798}

\bibitem[{{Kane} {et~al.}(2021{\natexlab{a}}){Kane}, {Li}, {Wolf}, {Ostberg},
  \& {Hill}}]{kane2021a}
{Kane}, S.~R., {Li}, Z., {Wolf}, E.~T., {Ostberg}, C., \& {Hill}, M.~L.
  2021{\natexlab{a}}, \aj, 161, 31, \dodoi{10.3847/1538-3881/abcbfd}

\bibitem[{{Kane} \& {Torres}(2017)}]{kane2017d}
{Kane}, S.~R., \& {Torres}, S.~M. 2017, \aj, 154, 204,
  \dodoi{10.3847/1538-3881/aa8fce}

\bibitem[{{Kane} {et~al.}(2020){Kane}, {Vervoort}, {Horner}, \&
  {Pozuelos}}]{kane2020e}
{Kane}, S.~R., {Vervoort}, P., {Horner}, J., \& {Pozuelos}, F.~J. 2020, \psj,
  1, 42, \dodoi{10.3847/PSJ/abae63}

\bibitem[{{Kane} {et~al.}(2021{\natexlab{b}}){Kane}, {Arney}, {Byrne}, {Dalba},
  {Desch}, {Horner}, {Izenberg}, {Mandt}, {Meadows}, \& {Quick}}]{kane2021d}
{Kane}, S.~R., {Arney}, G.~N., {Byrne}, P.~K., {et~al.} 2021{\natexlab{b}},
  Journal of Geophysical Research (Planets), 126, e06643,
  \dodoi{10.1002/jgre.v126.2}

\bibitem[{{Kinoshita}(1977)}]{kinoshita1977a}
{Kinoshita}, H. 1977, Celestial Mechanics, 15, 277, \dodoi{10.1007/BF01228425}

\bibitem[{{Kleine} {et~al.}(2009){Kleine}, {Touboul}, {Bourdon}, {Nimmo},
  {Mezger}, {Palme}, {Jacobsen}, {Yin}, \& {Halliday}}]{kleine2009}
{Kleine}, T., {Touboul}, M., {Bourdon}, B., {et~al.} 2009, \gca, 73, 5150,
  \dodoi{10.1016/j.gca.2008.11.047}

\bibitem[{{Lambrechts} \& {Johansen}(2014)}]{lambrechts2014b}
{Lambrechts}, M., \& {Johansen}, A. 2014, \aap, 572, A107,
  \dodoi{10.1051/0004-6361/201424343}

\bibitem[{{Laskar}(1990)}]{laskar1990}
{Laskar}, J. 1990, \icarus, 88, 266, \dodoi{10.1016/0019-1035(90)90084-M}

\bibitem[{{Laskar} {et~al.}(2011){Laskar}, {Fienga}, {Gastineau}, \&
  {Manche}}]{laskar2011a}
{Laskar}, J., {Fienga}, A., {Gastineau}, M., \& {Manche}, H. 2011, \aap, 532,
  A89, \dodoi{10.1051/0004-6361/201116836}

\bibitem[{{Laskar} {et~al.}(1993){Laskar}, {Joutel}, \&
  {Robutel}}]{laskar1993b}
{Laskar}, J., {Joutel}, F., \& {Robutel}, P. 1993, \nat, 361, 615,
  \dodoi{10.1038/361615a0}

\bibitem[{{Laskar} {et~al.}(2004){Laskar}, {Robutel}, {Joutel}, {Gastineau},
  {Correia}, \& {Levrard}}]{laskar2004c}
{Laskar}, J., {Robutel}, P., {Joutel}, F., {et~al.} 2004, \aap, 428, 261,
  \dodoi{10.1051/0004-6361:20041335}

\bibitem[{{Leandro} {et~al.}(2022){Leandro}, {Savian}, {Kochhann}, {Franco},
  {Coccioni}, {Frontalini}, {Gardin}, {Jovane}, {Figueiredo}, {Tedeschi},
  {Janikian}, {Almeida}, \& {Trindade}}]{leandro2022}
{Leandro}, C.~G., {Savian}, J.~F., {Kochhann}, M.~V.~L., {et~al.} 2022, Nature
  Communications, 13, 2941, \dodoi{10.1038/s41467-022-30075-3}

\bibitem[{{Lisiecki} \& {Raymo}(2005)}]{lisiecki2005}
{Lisiecki}, L.~E., \& {Raymo}, M.~E. 2005, Paleoceanography, 20, PA1003,
  \dodoi{10.1029/2004PA001071}

\bibitem[{{Lissauer} {et~al.}(2012){Lissauer}, {Barnes}, \&
  {Chambers}}]{lissauer2012a}
{Lissauer}, J.~J., {Barnes}, J.~W., \& {Chambers}, J.~E. 2012, \icarus, 217,
  77, \dodoi{10.1016/j.icarus.2011.10.013}

\bibitem[{{Lissauer} {et~al.}(2001){Lissauer}, {Quintana}, {Rivera}, \&
  {Duncan}}]{lissauer2001c}
{Lissauer}, J.~J., {Quintana}, E.~V., {Rivera}, E.~J., \& {Duncan}, M.~J. 2001,
  \icarus, 154, 449, \dodoi{10.1006/icar.2001.6692}

\bibitem[{{Lissauer} {et~al.}(2011){Lissauer}, {Ragozzine}, {Fabrycky},
  {Steffen}, {Ford}, {Jenkins}, {Shporer}, {Holman}, {Rowe}, {Quintana},
  {Batalha}, {Borucki}, {Bryson}, {Caldwell}, {Carter}, {Ciardi}, {Dunham},
  {Fortney}, {Gautier}, {Howell}, {Koch}, {Latham}, {Marcy}, {Morehead}, \&
  {Sasselov}}]{lissauer2011b}
{Lissauer}, J.~J., {Ragozzine}, D., {Fabrycky}, D.~C., {et~al.} 2011, \apjs,
  197, 8, \dodoi{10.1088/0067-0049/197/1/8}

\bibitem[{{Liu} {et~al.}(2020){Liu}, {Wu}, {Hinnov}, {Xi}, {He}, {Zhang}, \&
  {Yang}}]{liu2020c}
{Liu}, W., {Wu}, H., {Hinnov}, L.~A., {et~al.} 2020, Frontiers in Earth
  Science, 8, 178, \dodoi{10.3389/feart.2020.00178}

\bibitem[{{Ma} {et~al.}(2019){Ma}, {Meyers}, \& {Sageman}}]{ma2019c}
{Ma}, C., {Meyers}, S.~R., \& {Sageman}, B.~B. 2019, Earth and Planetary
  Science Letters, 513, 1, \dodoi{10.1016/j.epsl.2019.01.053}

\bibitem[{{Martin} \& {Livio}(2015)}]{martin2015b}
{Martin}, R.~G., \& {Livio}, M. 2015, \apj, 810, 105,
  \dodoi{10.1088/0004-637X/810/2/105}

\bibitem[{{Masset} \& {Snellgrove}(2001)}]{masset2001a}
{Masset}, F., \& {Snellgrove}, M. 2001, \mnras, 320, L55,
  \dodoi{10.1046/j.1365-8711.2001.04159.x}

\bibitem[{{Menou} \& {Tabachnik}(2003)}]{menou2003a}
{Menou}, K., \& {Tabachnik}, S. 2003, \apj, 583, 473, \dodoi{10.1086/345359}

\bibitem[{{Millholland} \& {Laughlin}(2019)}]{millholland2019a}
{Millholland}, S., \& {Laughlin}, G. 2019, Nature Astronomy, 3, 424,
  \dodoi{10.1038/s41550-019-0701-7}

\bibitem[{{Mogavero} \& {Laskar}(2021)}]{mogavero2021}
{Mogavero}, F., \& {Laskar}, J. 2021, \aap, 655, A1,
  \dodoi{10.1051/0004-6361/202141007}

\bibitem[{{Morbidelli} \& {Crida}(2007)}]{morbidelli2007d}
{Morbidelli}, A., \& {Crida}, A. 2007, \icarus, 191, 158,
  \dodoi{10.1016/j.icarus.2007.04.001}

\bibitem[{{Morbidelli} {et~al.}(2015){Morbidelli}, {Lambrechts}, {Jacobson}, \&
  {Bitsch}}]{morbidelli2015a}
{Morbidelli}, A., {Lambrechts}, M., {Jacobson}, S., \& {Bitsch}, B. 2015,
  \icarus, 258, 418, \dodoi{10.1016/j.icarus.2015.06.003}

\bibitem[{{Murray} \& {Dermott}(1999)}]{murray1999a}
{Murray}, C.~D., \& {Dermott}, S.~F. 1999, {Solar System Dynamics} ({Cambridge
  University Press}), \dodoi{10.1017/CBO9781139174817}

\bibitem[{{Olsen} {et~al.}(2019){Olsen}, {Laskar}, {Kent}, {Kinney},
  {Reynolds}, {Sha}, \& {Whiteside}}]{olsen2019}
{Olsen}, P.~E., {Laskar}, J., {Kent}, D.~V., {et~al.} 2019, Proceedings of the
  National Academy of Science, 116, 10664, \dodoi{10.1073/pnas.1813901116}

\bibitem[{{Park} {et~al.}(2021){Park}, {Folkner}, {Williams}, \&
  {Boggs}}]{park2021}
{Park}, R.~S., {Folkner}, W.~M., {Williams}, J.~G., \& {Boggs}, D.~H. 2021,
  \aj, 161, 105, \dodoi{10.3847/1538-3881/abd414}

\bibitem[{{Pierens} \& {Raymond}(2011)}]{pierens2011b}
{Pierens}, A., \& {Raymond}, S.~N. 2011, \aap, 533, A131,
  \dodoi{10.1051/0004-6361/201117451}

\bibitem[{{Quarles} {et~al.}(2020){Quarles}, {Barnes}, {Lissauer}, \&
  {Chambers}}]{quarles2020a}
{Quarles}, B., {Barnes}, J.~W., {Lissauer}, J.~J., \& {Chambers}, J. 2020,
  Astrobiology, 20, 73, \dodoi{10.1089/ast.2018.1932}

\bibitem[{{Quinn} {et~al.}(1991){Quinn}, {Tremaine}, \& {Duncan}}]{quinn1991}
{Quinn}, T.~R., {Tremaine}, S., \& {Duncan}, M. 1991, \aj, 101, 2287,
  \dodoi{10.1086/115850}

\bibitem[{{Raymond} {et~al.}(2024){Raymond}, {Kaib}, {Selsis}, \&
  {Bouy}}]{raymond2024a}
{Raymond}, S.~N., {Kaib}, N.~A., {Selsis}, F., \& {Bouy}, H. 2024, \mnras, 527,
  6126, \dodoi{10.1093/mnras/stad3604}

\bibitem[{{Raymond} {et~al.}(2009){Raymond}, {O'Brien}, {Morbidelli}, \&
  {Kaib}}]{raymond2009c}
{Raymond}, S.~N., {O'Brien}, D.~P., {Morbidelli}, A., \& {Kaib}, N.~A. 2009,
  \icarus, 203, 644, \dodoi{10.1016/j.icarus.2009.05.016}

\bibitem[{{Robertson} {et~al.}(2012){Robertson}, {Horner}, {Wittenmyer},
  {Endl}, {Cochran}, {MacQueen}, {Brugamyer}, {Simon}, {Barnes}, \&
  {Caldwell}}]{robertson2012b}
{Robertson}, P., {Horner}, J., {Wittenmyer}, R.~A., {et~al.} 2012, \apj, 754,
  50, \dodoi{10.1088/0004-637X/754/1/50}

\bibitem[{{Rosenthal} {et~al.}(2021){Rosenthal}, {Fulton}, {Hirsch},
  {Isaacson}, {Howard}, {Dedrick}, {Sherstyuk}, {Blunt}, {Petigura}, {Knutson},
  {Behmard}, {Chontos}, {Crepp}, {Crossfield}, {Dalba}, {Fischer}, {Henry},
  {Kane}, {Kosiarek}, {Marcy}, {Rubenzahl}, {Weiss}, \&
  {Wright}}]{rosenthal2021}
{Rosenthal}, L.~J., {Fulton}, B.~J., {Hirsch}, L.~A., {et~al.} 2021, \apjs,
  255, 8, \dodoi{10.3847/1538-4365/abe23c}

\bibitem[{{Saillenfest} {et~al.}(2019){Saillenfest}, {Laskar}, \&
  {Bou{\'e}}}]{saillenfest2019a}
{Saillenfest}, M., {Laskar}, J., \& {Bou{\'e}}, G. 2019, \aap, 623, A4,
  \dodoi{10.1051/0004-6361/201834344}

\bibitem[{{Spiegel} {et~al.}(2009){Spiegel}, {Menou}, \&
  {Scharf}}]{spiegel2009a}
{Spiegel}, D.~S., {Menou}, K., \& {Scharf}, C.~A. 2009, \apj, 691, 596,
  \dodoi{10.1088/0004-637X/691/1/596}

\bibitem[{{Spiegel} {et~al.}(2010){Spiegel}, {Raymond}, {Dressing}, {Scharf},
  \& {Mitchell}}]{spiegel2010b}
{Spiegel}, D.~S., {Raymond}, S.~N., {Dressing}, C.~D., {Scharf}, C.~A., \&
  {Mitchell}, J.~L. 2010, \apj, 721, 1308, \dodoi{10.1088/0004-637X/721/2/1308}

\bibitem[{{Vervoort} {et~al.}(2022){Vervoort}, {Horner}, {Kane}, {Kirtland
  Turner}, \& {Gilmore}}]{vervoort2022}
{Vervoort}, P., {Horner}, J., {Kane}, S.~R., {Kirtland Turner}, S., \&
  {Gilmore}, J.~B. 2022, \aj, 164, 130, \dodoi{10.3847/1538-3881/ac87fd}

\bibitem[{{Vervoort} {et~al.}(2024){Vervoort}, {Kirtland Turner}, {Rochholz},
  \& {Ridgwell}}]{vervoort2024}
{Vervoort}, P., {Kirtland Turner}, S., {Rochholz}, F., \& {Ridgwell}, A. 2024,
  Paleoceanography and Paleoclimatology, 39, e2023PA004826,
  \dodoi{10.1029/2023PA004826}

\bibitem[{{Walsh} {et~al.}(2011){Walsh}, {Morbidelli}, {Raymond}, {O'Brien}, \&
  {Mandell}}]{walsh2011c}
{Walsh}, K.~J., {Morbidelli}, A., {Raymond}, S.~N., {O'Brien}, D.~P., \&
  {Mandell}, A.~M. 2011, \nat, 475, 206, \dodoi{10.1038/nature10201}

\bibitem[{{Way} \& {Georgakarakos}(2017)}]{way2017a}
{Way}, M.~J., \& {Georgakarakos}, N. 2017, \apj, 835, L1,
  \dodoi{10.3847/2041-8213/835/1/L1}

\bibitem[{{Williams} \& {Kasting}(1997)}]{williams1997b}
{Williams}, D.~M., \& {Kasting}, J.~F. 1997, \icarus, 129, 254,
  \dodoi{10.1006/icar.1997.5759}

\bibitem[{{Williams} \& {Pollard}(2002)}]{williams2002}
{Williams}, D.~M., \& {Pollard}, D. 2002, International Journal of
  Astrobiology, 1, 61, \dodoi{10.1017/S1473550402001064}

\bibitem[{{Winn} \& {Fabrycky}(2015)}]{winn2015}
{Winn}, J.~N., \& {Fabrycky}, D.~C. 2015, \araa, 53, 409,
  \dodoi{10.1146/annurev-astro-082214-122246}

\bibitem[{{Wisdom}(2006)}]{wisdom2006b}
{Wisdom}, J. 2006, \aj, 131, 2294, \dodoi{10.1086/500829}

\bibitem[{{Wisdom} \& {Holman}(1991)}]{wisdom1991}
{Wisdom}, J., \& {Holman}, M. 1991, \aj, 102, 1528, \dodoi{10.1086/115978}

\bibitem[{{Wittenmyer} {et~al.}(2011){Wittenmyer}, {Tinney}, {O'Toole},
  {Jones}, {Butler}, {Carter}, \& {Bailey}}]{wittenmyer2011a}
{Wittenmyer}, R.~A., {Tinney}, C.~G., {O'Toole}, S.~J., {et~al.} 2011, \apj,
  727, 102, \dodoi{10.1088/0004-637X/727/2/102}

\bibitem[{{Wittenmyer} {et~al.}(2016){Wittenmyer}, {Butler}, {Tinney},
  {Horner}, {Carter}, {Wright}, {Jones}, {Bailey}, \&
  {O'Toole}}]{wittenmyer2016c}
{Wittenmyer}, R.~A., {Butler}, R.~P., {Tinney}, C.~G., {et~al.} 2016, \apj,
  819, 28, \dodoi{10.3847/0004-637X/819/1/28}

\bibitem[{{Wittenmyer} {et~al.}(2020){Wittenmyer}, {Wang}, {Horner}, {Butler},
  {Tinney}, {Carter}, {Wright}, {Jones}, {Bailey}, {O'Toole}, \&
  {Johns}}]{wittenmyer2020b}
{Wittenmyer}, R.~A., {Wang}, S., {Horner}, J., {et~al.} 2020, \mnras, 492, 377,
  \dodoi{10.1093/mnras/stz3436}

\bibitem[{{Wu} {et~al.}(2023){Wu}, {Hinnov}, {Zhang}, {Jiang}, {Yang}, {Li},
  {Xi}, {Ma}, \& {Wang}}]{wu2023e}
{Wu}, H., {Hinnov}, L.~A., {Zhang}, S., {et~al.} 2023, Geological Society of
  America Bulletin, 135, 712, \dodoi{10.1130/B36340.1}

\bibitem[{{Zachos} {et~al.}(2010){Zachos}, {McCarren}, {Murphy}, {R{\"o}hl}, \&
  {Westerhold}}]{zachos2010}
{Zachos}, J.~C., {McCarren}, H., {Murphy}, B., {R{\"o}hl}, U., \& {Westerhold},
  T. 2010, Earth and Planetary Science Letters, 299, 242,
  \dodoi{10.1016/j.epsl.2010.09.004}

\bibitem[{{Zeebe} \& {Lantink}(2024)}]{zeebe2024a}
{Zeebe}, R.~E., \& {Lantink}, M.~L. 2024, \aj, 167, 204,
  \dodoi{10.3847/1538-3881/ad32cf}

\bibitem[{{Zeebe} \& {Lourens}(2019)}]{zeebe2019b}
{Zeebe}, R.~E., \& {Lourens}, L.~J. 2019, Science, 365, 926,
  \dodoi{10.1126/science.aax0612}

\bibitem[{{Zhang} {et~al.}(2023){Zhang}, {Jin}, {Gillman}, {Liu}, {Wei}, {Li},
  \& {Zhang}}]{zhang2023k}
{Zhang}, R., {Jin}, Z., {Gillman}, M., {et~al.} 2023, Science China Earth
  Sciences, 66, 358, \dodoi{10.1007/s11430-021-9994-y}

\bibitem[{{Zhang} {et~al.}(2022){Zhang}, {Li}, {Fan}, {Da Silva}, {Shi}, {Gao},
  {Kuang}, {Liu}, {Gao}, \& {Li}}]{zhang2022e}
{Zhang}, T., {Li}, Y., {Fan}, T., {et~al.} 2022, Earth and Planetary Science
  Letters, 583, 117420, \dodoi{10.1016/j.epsl.2022.117420}

\bibitem[{{Zink} {et~al.}(2020){Zink}, {Batygin}, \& {Adams}}]{zink2020c}
{Zink}, J.~K., {Batygin}, K., \& {Adams}, F.~C. 2020, \aj, 160, 232,
  \dodoi{10.3847/1538-3881/abb8de}

\end{thebibliography}


\end{document}